\documentclass[prd,aps,twocolumn,nofootinbib]{revtex4}

\usepackage[utf8]{inputenc}
\usepackage{graphicx}
\usepackage{color}
\usepackage{hyperref}
\usepackage{xcolor}
\usepackage{amsmath}
\usepackage{multirow}
\usepackage{array}
\usepackage{float}
\usepackage{subcaption}
\hypersetup{
    colorlinks,
    linkcolor={red!80!black},
    citecolor={blue!80!black},
    urlcolor={green!50!black}
}

\def\gr{$\gamma$-ray}

\def\theta{\vartheta}

\def\be{\begin{equation}}
\def\ee{\end{equation}}
\def\ba{\begin{eqnarray}}
\def\ea{\end{eqnarray}}

\def\lsim{\raise0.3ex\hbox{$\;<$\kern-0.75em\raise-1.1ex\hbox{$\sim\;$}}}
\def\gsim{\raise0.3ex\hbox{$\;>$\kern-0.75em\raise-1.1ex\hbox{$\sim\;$}}}

\begin{document}

\title{Diffuse gamma-ray and neutrino emission
%flux 
from the Milky Way\\ and the local knee in the cosmic ray spectrum}
    \author{C.~Prévotat$^{1}$, Zh.~Zhu$^{2}$,  S.~Koldobskiy$^{3}$, A.~Neronov$^{2,4}$, D.~Semikoz$^{2}$ and  M.~Ahlers$^{5}$}

\affiliation{$^1$Sorbonne Université, CNRS, Institut d’Astrophysique de Paris, 98 bis boulevard Arago, 75014 Paris, France}
\affiliation{$^2$Université de Paris Cite, CNRS, Astroparticule et Cosmologie, F-75013 Paris, France}
\affiliation{$^{3}$Sodankyl\"a Geophysical Observatory and Space Physics and Astronomy Research Unit, University of Oulu, 90014 Oulu, Finland}
\affiliation{$^4$Laboratory of Astrophysics, Ecole Polytechnique Federale de Lausanne, CH-1015, Lausanne, Switzerland}
\affiliation{$^5$Niels Bohr Institute, University of Copenhagen, Blegdamsvej 17, 2100 Copenhagen, Denmark}
\begin{abstract}
The LHAASO observatory has recently measured 
%the 
details of the cosmic-ray (CR) spectrum in the knee region (1 -- 10 PeV) with unprecedented precision, including its average CR mass composition and the spectrum of the proton component. We use these precision measurements, combined with direct measurements of CRs by space-based detectors, to derive predictions for the spectrum of diffuse $\gamma$-ray and neutrino emission from the interstellar medium under the assumption that the CR spectrum is universal throughout the Milky Way. We compare these predictions with the Fermi-LAT and LHAASO measurements of the diffuse \gr\ flux from inner and outer Galactic Plane regions and with estimates of the neutrino flux based on the IceCube data for the same Galactic Plane regions. 
%In this work we compare predictions of gamma-ray production of those models of cosmic ray distribution in the Galaxy  with line of sight distribution of gamma-rays with data of Fermi LAT and LHAASO. We describe secondary gamma-ray and neutrino production with AAFrag code. 
We notice that the model predictions exceed LHAASO \gr\ measurements at energies above 100~TeV. 
This excess can be interpreted within 
a CR knee model assuming a ``local PeV CR bubble''.
Within this model, we infer the  extension  of the local PeV CR bubble of $1.5\pm 0.3$~kpc in the direction of the inner Galaxy and of $0.7\pm 0.3$~kpc toward the outer Galaxy. 
%{\bf In the inner Galaxy extension is combination local bubble and other similar bubbles integrated along line of sight.  Our model predicts large variations of cosmic ray flux along Galactic plane due to presence/absence of  CR bubbles in given part of Galaxy.}
\end{abstract}
\maketitle

%%%%%%%%%%%%%%%%%%%%%%%%%%%%%%%%%%%%%%%%%%%%%
\section{Introduction}
\label{Sec:Intro}

The cosmic ray (CR) spectrum 
has been 
measured in the Solar System from sub-GeV energies to the highest energies $E>10^{20}$ eV~\cite{1984acr..book.....B,Kachelriess:2019oqu}. 
CRs with energies $E<100$ TeV are observed directly with 
{balloon-borne of space-based detectors.} %detectors in  upper atmosphere or in space around Earth. 
{The} AMS-02 detector provides {as yet} the highest precision measurements of spectra of individual nuclei up to  TeV  energies~\cite{AMS02_light, AMS02_Mg, AMS02_Al, AMS02_Fe}. {The} CALET~\cite{CALET:2023nif} and DAMPE~\cite{DAMPE_protons, DAMPE_CNO, DAMPE_He} detectors provide direct measurements of the spectra of individual {CR} nuclei up to 100~TeV. Above 100 TeV, measurements of the CR flux are provided by ground-based experiments. Most recently, {the} LHAASO detector made  precise  measurements of the total spectrum, 
the average logarithmic mass ({$\langle\ln A\rangle$}) spectrum~\cite{LHAASO:2024knt} and of the proton spectrum~\cite{LHAASO:2025byy}. 

The CR spectrum has several features, which can be characterized by {a} change {$\Delta\alpha$ of the slope} of its power-law {spectrum $dN/dE \propto E^{-\alpha}$}. The most famous feature is {the} {\it knee} at the energy $E\simeq 4$~PeV where the spectral slope  changes by $\Delta \alpha\simeq 0.4$~\cite{knee_discovery,Hoerandel:2004gv,Kachelriess:2019oqu}. 
Recent LHAASO measurements find the proton knee  at the energy 3 PeV with a change of the spectral slope by {$\Delta\Gamma\simeq 1.0$} preceeded by an unexpected hardening  around PeV energy~\cite{LHAASO:2025byy}.

Measurements of {the} primary-to-secondary CR flux ratios  indicate that most CRs escape from the Galactic environment via diffusion before interacting with the interstellar medium.  This imposes a constraint on the Galactic magnetic field, which has to have {a} non-zero component in {the} direction perpendicular to {the} Galactic Plane~\cite{Giacinti:2017dgt}. Measurements of the slopes of the 
{primary-to-secondary CR} %primary and secondary cosmic ray 
species provide information on 
{the rigidity-dependent strength of CR diffusion} %details of CR propagation 
through the Galactic magnetic field. A generic expectation of CR propagation models is that in the energy range in which interactions of the primary nuclei do not significantly influence their spectra,  all nuclei should have approximately {the} same spectrum as protons since both acceleration to relativistic energies and 
{diffusion} %propagation 
are sensitive only to {the} rigidity of nuclei 
{${\cal R}=pc/Ze \simeq E/Ze$} %${\cal R}=E/Z$ 
(energy $E$ over {charge $Ze$)}~\cite{1984acr..book.....B}. % atomic number $Z$)
However, this expectation contradicts the observational data and {a} number of models overcoming this difficulty {have} been proposed~\cite{2013ApJ...763...47P,Kachelriess:2018ser,Evoli:2021ugn}. These models can be distinguished using complementary \gr\ and neutrino data
{in comparison to the predicted diffuse emission} %\gr s and neutrinos are 
produced in the course of CR propagation through the interstellar medium. Hence, \gr\  and neutrino spectra provide information on the CR spectrum at different locations across the Galactic Disk.

Measurements of the spectra of diffuse emission from the Galactic {Plane} %Disk 
are now available over a broad energy range from sub-GeV up to PeV due to Fermi-LAT~\cite{2012ApJ...750....3A,Fermi-LAT:2016zaq,Neronov:2019ncc},
{Tibet-AS$\gamma$~\cite{TibetASgamma:2021tpz}, HAWC~\cite{HAWC:2023wdq},}
and LHAASO~\cite{LHAASO:2023gne}. 
In what follows we use 
{measurements from Fermi-LAT and LHAASO} %these measurements 
to test 
%the 
models of {the} Galactic CR population.  
%Also some hints of neutrino signal from Milky Way were independently seen by ANTARES \cite{ANTARES:2022izu} and Baikal-GVD \cite{Baikal-GVD:2024kfx}. 
Neutrino emission %signal 
from {our Galaxy was recently} observed by IceCube with {a} significance above 4$\sigma$~\cite{Abbasi2023}. {Additional} hints of a neutrino signal from {the} Milky Way were independently seen by ANTARES~\cite{ANTARES:2022izu} and Baikal-GVD~\cite{Baikal-GVD:2024kfx}. In this work we use 
%templates from 
{the result of the IceCube analysis~\cite{Abbasi2023} re-scaled to the inner and outer Galaxy studied by LHAASO} in order to compare {the} level of neutrino flux in {the} Milky Way with model predictions.

Specifically, we compare predictions of two models of 
Galactic CRs. In Model A we assume that breaks of {the} CR spectrum up to {the} knee are due to contributions of several different populations of sources, using {the} so-called Peters cycles~\cite{Peters61, Zatsepin62}. 
{In this model, the observed} spectra of all  
nuclei {corresponding to individual source populations} have the same shape as a function of rigidity. The abundances of different nuclei are allowed to vary. For Model B we assume that all observed breaks in CR proton spectrum are propagation features. {Following Ref.~\cite{Schwefer_2023}, we} assume 
%proton spectrum at sources 
{a simple power-law spectrum of nuclei at the sources with variable normalization}
and use {the} DRAGON-2 code~\cite{Evoli_2017} to 
{derive the stationary solution of CRs} %model propagation of cosmic rays 
in the Galaxy. We fit both models to CR data to find the best fit parameters of the models and check their (in)consistency with the CR data.
% \MA{\bf (This last sentence sounds like a repetition. Also, do we really fully fit model B or are we not simply using the best-fit diffusion coefficient of Mertsch et al.?)}

Following {the} approach of Ref.~\cite{Prevotat:2024ynu},  we calculate the CR distribution in the Galaxy for both models and use this distribution to compute the diffuse \gr\ and neutrino emission from the Galactic {Plane} % disk
taking into account interactions of CRs in {the interstellar medium} %Galaxy with gas, 
and absorption of $\gamma$-rays {before arriving at} %on the way to 
Earth. We compare predictions of both models 
{to} % with a combination of 
Fermi-LAT and LHAASO data. 

The paper is organized as following: Sections \ref{Sec:CR_fit} describes details of the CR Models A and B.  In Section \ref{Sec:diffuse_data} we compute diffuse \gr\ and neutrino fluxes from Galactic Disk expected with the CR distributions of Models A and B. In Section \ref{sec:results} we compare model predictions with the data and in section \ref{Discussion} we discuss  the results.

%%%%%%%%%%%%%%%%%%%%%%%%%%%%%%%%%%%%%%%%%%%%%%
\section{Cosmic ray population models}
\label{Sec:CR_fit}

We consider two alternative models of CR distribution across the Galactic Plane. %Disk. 
For each model, we fit the predicted CR spectra
to the measurements of CR % the spectra of cosmic ray 
nuclei in four elemental groups (protons, helium, intermediate and heavy) 
measured by AMS-02~\cite{AMS02_light, AMS02_Mg, AMS02_Al, AMS02_Fe}, DAMPE~\cite{DAMPE_protons, DAMPE_CNO, DAMPE_He}, LHAASO~\cite{LHAASO:2024knt}, and Auger~\cite{Auger_all, Auger_fractions}. We also include the recently published proton 
spectral data by LHAASO~\citep{LHAASO:2025byy}. To compensate for the absence of measurement of heavier elements by LHAASO, we use the all-particle spectrum and mean logarithmic mass of the CRs as measured by LHAASO~\cite{LHAASO:2024knt}. To account for the effects of solar modulation in the low-energy CR data we employ
local interstellar spectra from Ref.~\cite{Boschini2020} which agree with AMS-02 data inside the heliosphere and Voyager observations outside. For aluminum and iron, we 
use the spectra presented in Refs.~\cite{Al_LIS} and \cite{Fe_LIS}, which account for the first measurements of these elements by AMS-02.

%%%%%%%%%%%%%%%%%%%%%%%%%%%%%%%%%%%%%%%%%%%%%%
\subsection{Cosmic ray spectrum fit using Peters cycles}
%\label{Sec:CR_fit}

\begin{figure*}
    \centering
    \includegraphics[width=1\linewidth]{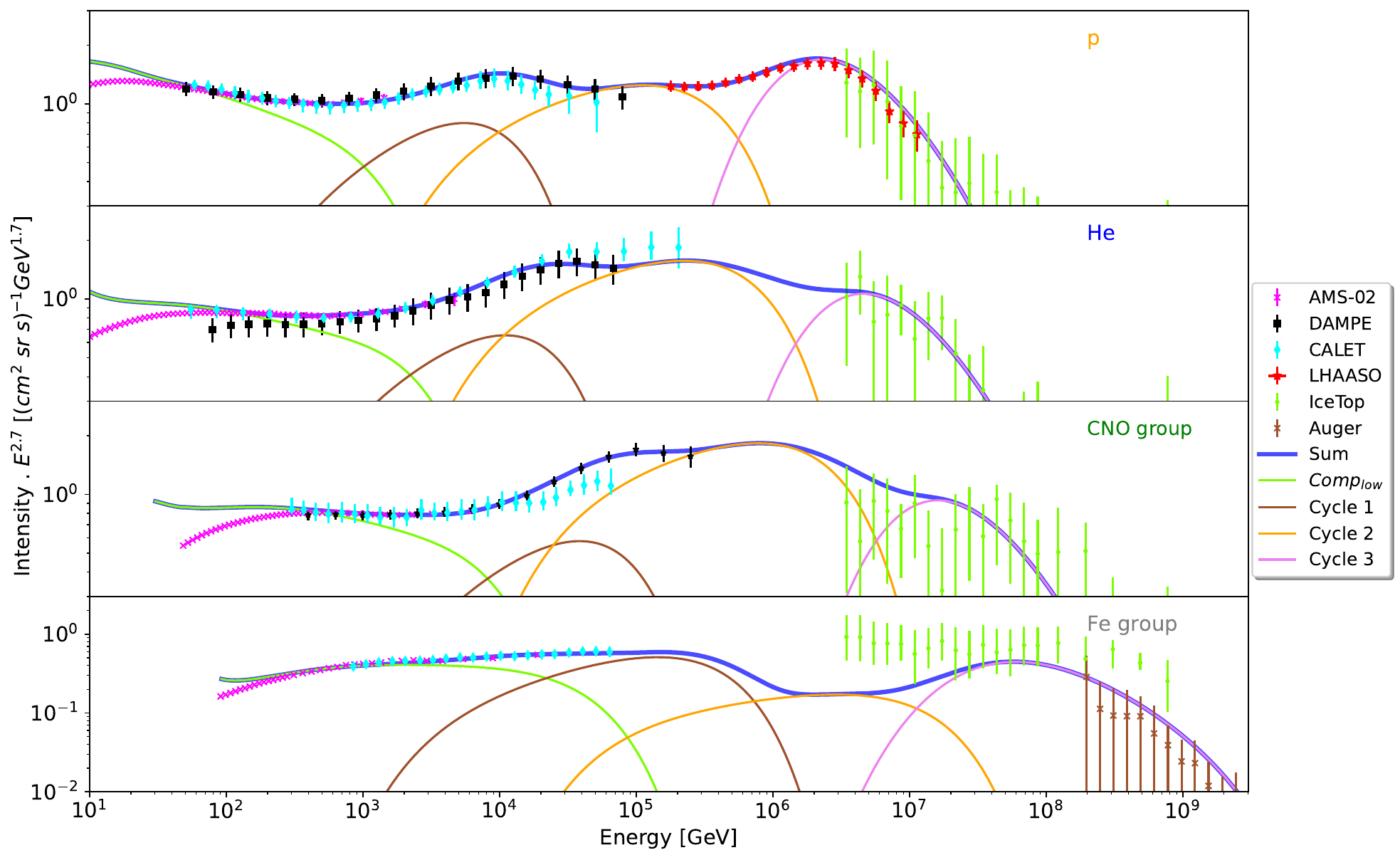}
    \caption{Cosmic ray spectra obtained for p, He, CNO and heavy elemental groups with Model A, composed of three Peters cycles and an additional low energy component. The data are from AMS-02~\cite{AMS02_light, AMS02_Mg, AMS02_Al, AMS02_Fe}, DAMPE~\cite{DAMPE_protons, DAMPE_CNO, DAMPE_He}, CALET~\cite{CALET_protons, CALET_He, CALET_Heavy, CALET_Fe}, LHAASO~\cite{LHAASO:2024knt}, IceTop~\cite{Aartsen2019} and Auger~\cite{Auger_all, Auger_fractions}. }
    \label{fig:peter_cycles}
\end{figure*}

\begin{figure*}
    \centering
    \includegraphics[width=1\linewidth]{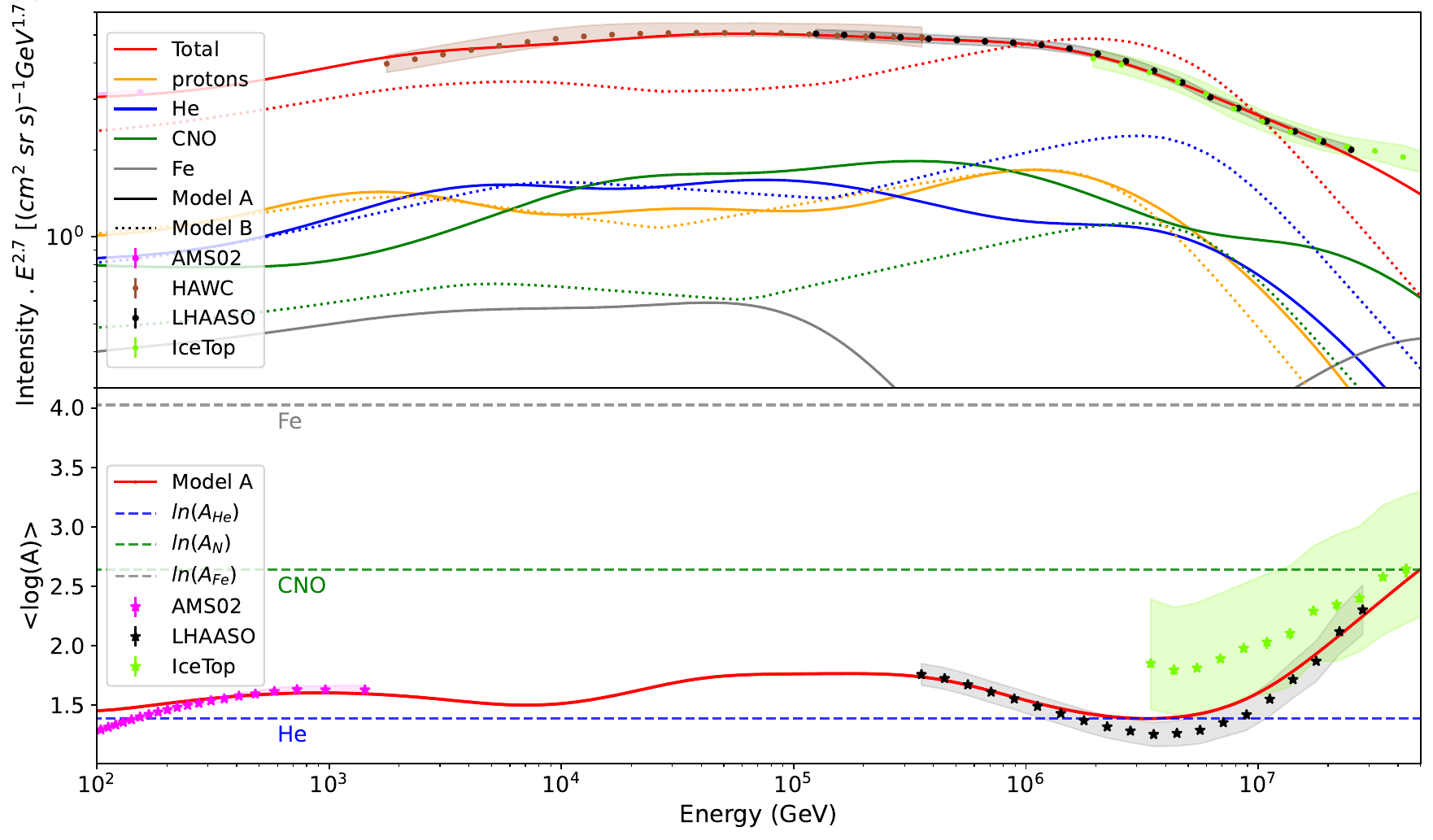}
    \caption{Result of the fit for the all-particles spectrum, and mean logarithmic mass of the CR spectrum with model A. The data are from AMS-02~\cite{AMS_Proton, AMS_He, AMS02_Al, AMS02_Mg, AMS02_Fe}, HAWC~\cite{HAWC_all}, LHAASO~\cite{LHAASO:2023gne} and IceTop~\cite{data_Icecube}. Only LHAASO’s data points were used in the fit. }
    \label{fig:all_particles+lnA}
\end{figure*}

Our Model A uses Peters cycles  
for modeling of Galactic CR population~\cite{Peters61, Zatsepin62}.   The rigidity spectra of different elemental groups only differ in their relative normalisations\footnote{This assumption is valid for primary CR nuclei in the energy range in which spallation reactions do not lead to significant changes in the spectrum on the time scale of residence of CRs in the Galaxy}. The method is very similar to what is presented in Ref.~\cite{Prevotat:2024ynu}. For the first two cycles (index $i$) we choose as templates power-law spectra with low- and high-energy exponential cutoffs ${\cal R}_{i0}$ and ${\cal R}_{i1}$, respectively, for each elemental group (index $j$) in the form: 
\be
I_{i,j} = \frac{\mathcal{N}_{i,j}}{(eZ_j {\cal R}_{i0})^{2.5}} \left(\frac{\cal R}{{\cal R}_{i0}}\right)^{-\alpha_i}  e^{{-{{\cal R}_{i0}}/{\cal R}}-{{\cal R}/{{\cal R}_{i1}}}}\,.
\label{eq:power_law}
\ee

Here, $\cal R$ is the rigidity in GV, $Z_j$ the atomic number of the elemental group $j$ (1, 2, 7 or 26), $\mathcal{N}_{i,j}$ is a normalisation constant, different for each element $j$ and population $i$, and $\alpha_i$ is the 
power-law index. % index of the power law  and $E_{i0,j}$ ($E_{i1,j}$) are energy cut-offs. 
To better fit LHAASO’s constraining data, the last ($i=3$) 
cycle is modeled with a broken power-law, multiplying Eq.~\ref{eq:power_law} by a factor 
\be
\eta({\cal R}) \equiv \left(1 + \left(\frac{\cal R}{{\cal R}_b}\right)^{\delta}\right)^{\frac{\alpha_4 - \alpha_3}{\delta}}
\ee

where ${\cal R}_b$ is the energy of the break, and $\delta$ characterizes the sharpness between the two regimes. 

An additional parameter is introduced to ensure a smooth transition in the region where we corrected for solar modulations, described by analytical formulae introduced in Refs.~\cite{Boschini2020, Al_LIS, Fe_LIS} and the first Peters cycle: we multiplied the solar modulated component by an exponential high-energy cut-off, with parameter ${\cal_R}_{01}$.

The fitting procedure resulted in a reduced chi-square of 0.34, results of the fit are showed in Figs.~\ref{fig:peter_cycles} and \ref{fig:all_particles+lnA}. We 
compare to data by CALET and LHAASO; while CALET differs from DAMPE at the highest energy points, it was shown in Ref.~\cite{Prevotat:2024ynu} that the difference in the induced $\gamma$-ray and neutrino emissions are small. Also, the result of our fit is found to match IceTop data for protons and helium; on the other hand, the fit results in a higher intensity in the CNO group, and a lower one in the heavy elements group, consistent with the higher values of $\langle\ln A\rangle$ found by IceTop compared to LHAASO.  A noticeable feature is that the highest energy part of LHAASO’s data corresponding to the knee region is described by our third Peters cycle, which appears only to fit this region of the CR spectrum.

This fit is only representing the local CR spectrum; in order to account for the spatial distribution of CRs, we follow the simple model of Ref.~\cite{Lipari:2018gzn} and assume that the radial distribution of CR sources follows pulsar distribution in the Galactic Disk and assume that the global CR spectral density confined to the Galactic diffusion halo can be approximated by that inferred from the local spectrum. The resulting fluxes of $\gamma$-ray and neutrinos are practically indistinguishable from a more accurate (but more time-consuming method) that accounts for approximate source distributions in the Galactic Plane and diffusive escape from the Galaxy.

%%%%%%%%%%%%%%%%%%%%%%%%%%%%%%%%%%%%%%%%%%%%%%%%%%%%%%%%
\subsection{Cosmic Ray Spectrum Fit Using Diffusion Break}
%\label{Sec:DRAGON}

For Model B, we 
{account for spectral features in the local CR spectrum via} % introduce the 
spectral changes that arise from the diffusive propagation of CRs in the Galactic magnetic field. We implement five breaks in the power-law rigidity dependence of diffusion coefficient $D(\mathcal{R})$ adopted from Ref.~\cite{Schwefer_2023},
\begin{align}
\label{eq: diffusion 5break}
D(\mathcal{R}) & =D_0 \beta\left(\frac{\mathcal{R}}{\mathcal{R}_{12}}\right)^{-\delta_1} \notag \\
& \times \prod_{i=1}^5\left(1+\left(\frac{\mathcal{R}}{\mathcal{R}_{i(i+1)}}\right)^{1 / s_{i(i+1)}}\right)^{-s_{i(i+1)}\left(\delta_{i+1}-\delta_i\right)}.
\end{align}
Here, $D_0$ is the normalization constant, $\beta$ the CR velocity in units of the speed of light, $\delta_i$ the power-law index in six adjacent rigidity ranges, $\mathcal{R}_{i(i+1)}$ the rigidity at the transition $\delta_i \to \delta_{i+1}$ and $S_{i(i+1)}$ parametrizes the softness of this transition. We implement this diffusion model into the DRAGON-2 code~\cite{Evoli_2017}, which numerically solves the full CR transport equations. 
%The result cosmic ray flux is made to fit the data from AMS \cite{AMS_Proton}, DAMPE \cite{DAMPE_protons}, and LHAASO \cite{LHAASO:2025byy}.  
The details of the simulation, including the parameter values and model set-up, are provided in the Appendix~\ref{Appendix:dragon}.

%%%%%%%%%%%%%%%%%%%%%%%%%%%%%%%%%%%%%%%%%%%%%%
\section{Gamma-ray and neutrino flux from inner and outer Galaxy}
\label{Sec:diffuse_data}

\begin{figure*}
    \centering
    \includegraphics[width=1\linewidth]{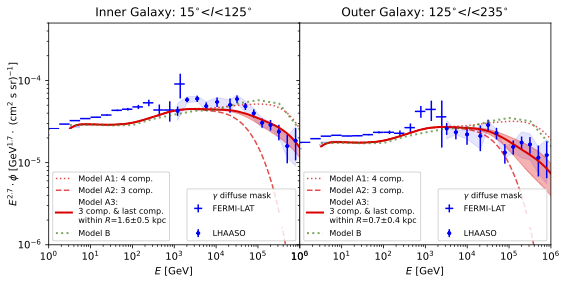}
    \caption{{The} \gr\ flux for Inner and Outer Galaxy regions measured by LHAASO (with the source mask). Crosses represent Fermi-LAT data and squares are for LHAASO. Light blue color shows the expected $\pi^0$ Fermi-LAT flux as discussed in the text. Red lines denote Model~A with several variations (see the text for details), while green dotted line shows result of Model~B simulations. }
    \label{fig:gamma_flux}
\end{figure*}

\begin{figure*}
    \centering
    \includegraphics[width=1\linewidth]{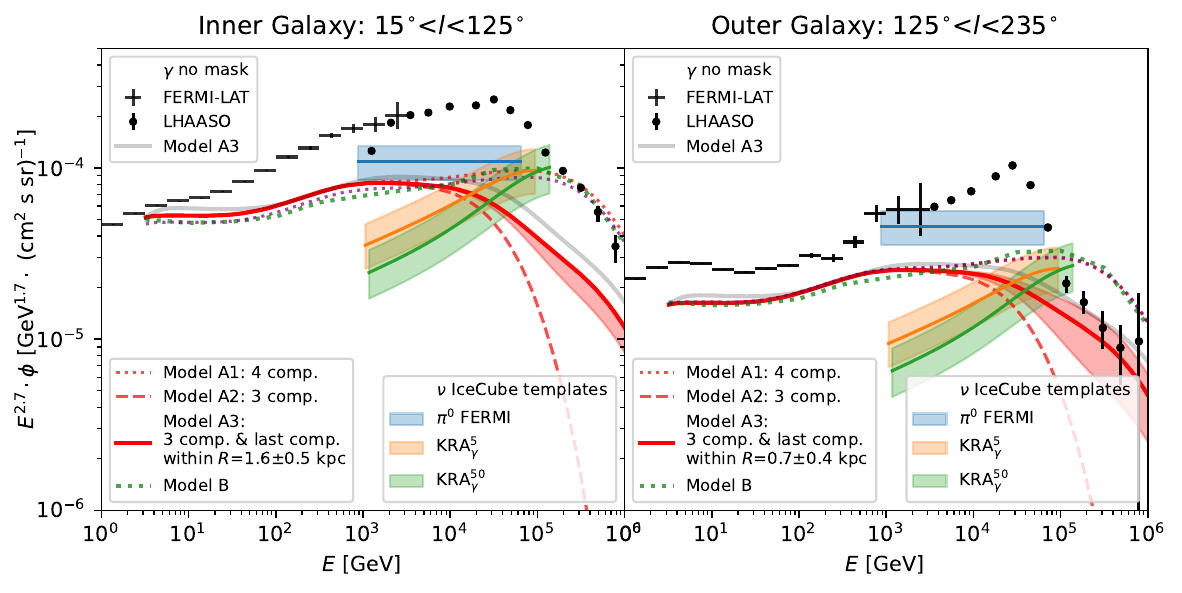}
    \caption{All flavor neutrino flux for Inner and Outer Galaxy regions as measured by LHAASO (without the source mask). Comparison between the models (color and line coding follows Fig.~\ref{fig:gamma_flux}) and three model-dependent data-based estimations by IceCube collaboration~\cite{Abbasi2023}. In addition, with purple dotted line we show the variation of 4-component Model A, where the different CR density gradient~\cite{Blasi:2011fi} was taken. Black crosses and squares show the unmasked \gr\-ray flux measured by Fermi-LAT and LHAASO, respectively. Gray lines show the expected diffuse \gr\ flux as calculated by Model A3.}
    \label{fig:nu_flux}
\end{figure*}

We use Models A and B to calculate {the} diffuse \gr\ and neutrino flux from the Galactic {Plane}. % Disk.
We use the AAfrag code~\cite{Kachelriess2019a,Koldobskiy2021b} to compute radiative yields of CR interactions with the interstellar gas. 
For the gas distribution in the Galactic disk we use atomic (H$_\mathrm{I}$) and molecular (H$_2$) hydrogen component models from Ref.~\cite{Lipari:2018gzn}.
The gradients of {the} CR density for Model A were 
{accounted for using the approximations of} %taken according to 
Refs.~\cite{Lipari:2018gzn} and \cite{Blasi:2011fi}, while for Model B Ref.~\cite{Ferri_re_2001} was used. 
% \MA{\bf (I think we are here mixing up the distribution of CRs with that of gas in Ref.~\cite{Ferri_re_2001}. Please check!) %% Checked, same paper cited and used by DRAGON code and also in Mertsch's paper %%.

{In all cases, the diffuse emission in $\gamma$-rays and neutrinos is determined by line-of-sight intergrals} %We perform line-of-sight integration 
of the yields of CR interactions with the interstellar gas, taking into account 
{$\gamma$-ray absorption} % as $\gamma$-rays 
by {the} cosmic microwave background and interstellar radiation fields, following Ref.~\cite{Celli_2017} and Ref.~\cite{Vernetto2016}. 

Figure~\ref{fig:gamma_flux} shows a comparison of predictions of the two CR models with observational data of Fermi-LAT and LHAASO for inner ($155^\circ<l<125^\circ$) and outer 
%Galaxy 
($125^\circ<l<235^\circ$) regions of the Galactic Plane ($|b|<5^\circ$). The \gr\ data shown in the plots are from 
LHAASO~\cite{Cao2025} with gray squares indicating the total flux and blue squares representing the diffuse flux corresponding to the sky region with LHAASO sources masked as described in Ref.~\cite{Cao2025}. LHAASO data are complemented with Fermi-LAT data for exactly the same sky regions, also showing total flux and the diffuse flux corresponding to the flux from regions masked as described in Ref.~\cite{Cao2025}. For our analysis 
%of Fermi-LAT data
we use publicly available {Fermi-LAT} data 
{collected} %in the time range 
between September 2008 and March 2025.  We select events from the P8R3\_SOURCEVETO\_V3 class
{and} %. We 
estimate exposure toward a given sky direction using a chain of {\it gtltcube-gtexpcube2} routines of {the} Fermi Science Tools\footnote{https://fermi.gsfc.nasa.gov/ssc/data/analysis/software/}. The total flux from a sky region is calculated by summing fluxes on a grid of sky directions covering 
%the regions of interest 
$15^\circ<l<125^\circ$ {and} $|b|<5^\circ$ for the inner Galaxy and $125^\circ<l<235^\circ$ {and} $|b|<5^\circ$ for the outer Galaxy for the total flux and for the parts of these regions with the LHAASO sources masked for the diffuse flux. %\MA{\bf (I do not understand this last sentence. Is this related to the ``$\pi_0$ (0.84)'' component in Figure~\ref{fig:gamma_flux}? Does this only affect the first three/four bins in Figure~\ref{fig:gamma_flux}? The choice of colours in Figure~\ref{fig:gamma_flux} could be improved.)}

The shape of the 
%model 
\gr\ spectra computed based on Models 
{A and B} % 1 and 2
described in the previous section is fixed by the shape of the CR spectrum. Red and green dotted lines in Fig.~\ref{fig:gamma_flux} show 
%the spectra of 
diffuse \gr\ 
{spectra} %flux 
for the two models. One can see that the two 
{model predictions} %shapes 
agree with each other, up to small discrepancies related to slightly different procedures for fitting of the CR flux in the approaches of Model A and B. We consider these difference as a model-related systematic error, {at the 10\% level.} % \sim 10\%$ level.

The spectra of diffuse \gr\ flux are normalized at 10~GeV energy according to the 
%Fermi-LAT 
detailed modeling of the 
{Fermi-LAT data on} diffuse emission in {the GeV energy} range~\cite{2012ApJ...750....3A,Fermi-LAT:2016zaq}. In this energy range, neutral pion decay 
%\gr\ emission 
provides the dominant contribution to the total diffuse \gr\ flux, but part of emission is from CR electrons, produced via inverse Compton and bremsstrahlung.
%mechanisms. 
The contribution from electrons can be distinguished from {that of} pion decay 
%emission because it is differently distributed on
{from the angular distribution in} the sky. We use the result of Ref.~\cite{Fermi-LAT:2016zaq} that has found that the pion decay component constitutes $0.85$ of the total diffuse flux, and emission from electrons contributes only $0.15$ of the diffuse flux at low Galactic latitude $|b|<10^\circ$. In our analysis this is perhaps a lower bound on the top of the  pion decay component fraction, because we are interested in a region at still lower Galactic latitude $|b|<5^\circ$. We consider the uncertainty of normalisation of the diffuse flux at 10~GeV as still another modeling uncertainty, at the level of {10\%.} %\sim 10\%$. 

Figure~\ref{fig:nu_flux} shows the expected diffuse all-flavor neutrino flux that accompanies the diffuse \gr\ flux from the same regions of the Galactic Plane. The neutrinos are produced together with \gr s in decays of {charged} pions and the shape and normalization of neutrino spectrum is calculated simultaneously with the shape for the \gr\ spectrum within the AAfrag code. We compare the predictions of the neutrino flux
{to that observed by IceCube~\cite{Abbasi2023} by rescaling their all-sky angular-averaged diffuse flux fits based on three different emission models (Fermi $\pi^0$ (blue), KRA${}_\gamma^5$ (orange), and KRA${}_\gamma^{50}$ (green)) to the corresponding angular regions along the Galactic Plane.} %from the two sky regions of interest with estimates of the neutrino flux from these regions based on the Galactic diffuse neutrino flux templates used by IceCube collaboration in the analysis of the signal from the Galactic Plane \cite{Abbasi2023}. 
{Note that the overall normalization of the Galactic neutrino signal in Ref.~\cite{Abbasi2023} is found by fitting all-sky IceCube data to the three corresponding angular templates assuming a fixed spectrum with variable overall normalization. The spectral features indicated by IceCube's observation are therefore representing an all-sky average, rather than the actual spectrum in the inner and outer Galaxy.} %from an all-sky fitting of IceCube data and the uncertainty of IceCube template normalization
{Nevertheless, the overall level of the rescaled fluxes allows for a comparison to the flux level of our model predictions in the two Galactic Plane regions while the spread between different fit templates} reflects uncertainty of the all-sky model, rather than uncertainty of the flux from the two regions of interest in our analysis. 
Reassuringly, {the} neutrino flux expectations from Models A and B agree with the IceCube model template flux estimates in the 100~TeV range, except for the {Fermi $\pi^0$} template in the outer Galaxy region. 
%\MA{\bf (I abbreviated this paragraph. We should not read too much into IceCube's observation rescaled to our two regions. Soon, there will be better data to compare to.)}

%%%%%%%%%%%%%%%%%%%%%%%%%%%%%%%%%%%%%%%%%%%%%%
\section{Model-data comparison}
\label{sec:results}

{The} comparison of the predictions of Models A and B with \gr\ data in Fig.~\ref{fig:gamma_flux} shows that the expected pion decay  flux is below the measurements of the diffuse flux by Fermi-LAT and LHAASO in the energy range below approximately 30~TeV.  In the energy range 1-30~TeV, the model flux expectations are consistent with LHAASO and Fermi-LAT data, so that the pion decay component can in principle constitute 100\% of the diffuse flux, with the diffuse flux originating from inverse Compton scattering by CR electrons vanishing in this energy range. This is expected because the CR electron spectrum softens in {the} multi-TeV energy range~\cite{DAMPE:2017fbg} and because inverse Compton scattering by multi-TeV electrons proceeds in Klein-Nishina regime with scattering cross-section decreasing with energy.

The pion decay model flux exceeds the diffuse emission flux measured by LHAASO in the inner Galaxy above 30~TeV. In the outer Galaxy, the pion model flux exceeds LHAASO diffuse \gr\ flux measurements at the energies above 100~TeV. This fact has already been noticed in Ref.~\cite{Prevotat:2024ynu} based on the modeling that did not include the LHAASO proton flux data. One can see that the inclusion of precision measurement of the proton spectrum by LHAASO confirms the conclusion of~\cite{Prevotat:2024ynu}. 

Comparing the predictions of the pion decay model with the neutrino flux estimates from IceCube (Fig.~\ref{fig:nu_flux}) shows that our analysis agrees well with 
{the expected flux based on the rescaled neutrino flux level} % estimates of the neutrino flux 
of Ref.~\cite{Abbasi2023} in the 100~TeV for all the three model templates.
%considered in Ref. \cite{Abbasi2023}. 
The {Fermi $\pi^0$} template 
%model 
provides a flux estimate that is a factor of two above our model flux in the outer Galaxy, but the uncertainty of the IceCube template flux estimate is large and, as mentioned above, the uncertainty band shown is that of the overall sky flux normalization, which is an under-estimate of uncertainty of the flux from the $125^\circ<l<250^\circ$ \ $|b|<5^\circ$ region. Figure \ref{fig:nu_flux} also shows the total \gr\ flux measurement of Fermi-LAT and LHAASO. We show this measurement for comparison, to highlight the fact that the {Fermi $\pi^0$} template neutrino flux estimate is at the level of total \gr\ flux at 100~TeV. The total flux includes the flux of resolved sources, along with the diffuse emission flux and hence it is (by definition) an upper bound on the diffuse \gr\ flux. Given that the neutrino and \gr\ pion decay fluxes are comparable, it can also be considered as an upper bound on the neutrino flux. One can see that the {Fermi $\pi^0$} template %model 
flux is at the level of the total \gr\ flux  in the outer Galaxy. %\MA{\bf (Are we comparing directly $\gamma$-rays to neutrinos? There are $\mathcal{O}(1)$ factors that might play a role.)} 
Moreover, if extrapolated to lower energies with the same 2.7 spectral slope (as horizontal line in Fig. \ref{fig:nu_flux}) the model flux estimate will largely exceed the total \gr\ flux upper bound in the energy range below $\sim 300$~GeV. The KRA$_\gamma$ template flux estimates agree with our model flux estimate at 100~TeV but disagree in the TeV range, by construction. In our model, the CR spectrum everywhere in the Galactic disk is assumed to be identical to the locally measured CR spectrum that has an average slope close to 2.7 over a broad energy range. To the contrary, in the KRA$_\gamma$ models the CR spectrum is harder, with the slope 2.5 instead of 2.7. Because of this slope difference, our model can agree with the KRA$_\gamma$ models only at one fixed energy. 

\section{Discussion}
\label{Discussion}

Comparison of the model predictions for the diffuse \gr\ and neutrino fluxes based on the LHAASO precision data on the CR spectrum in the knee region shown in Figs.~\ref{fig:gamma_flux} and \ref{fig:nu_flux} shows that these predictions are consistent with Fermi-LAT and LHAASO \gr\ data on diffuse emission from the Galactic disk in both inner and outer Galaxy up to 30~TeV energy.

The pion decay model predictions are somewhat below the diffuse flux measurements in 1-30~TeV band in the inner Galaxy. It is not immediately possible to draw robust conclusions on the origin of this discrepancy, because of large uncertainties of both the atomic nuclei and electron CR spectra in the inner part of the Milky Way. {The} electron spectrum is generically expected to be variable across the Galactic 
{Plane} %Disk 
because of the short cooling time of electrons with respect to synchrotron and inverse Compton energy losses. Even though spectra of protons and nuclei are expected to experience smaller variations, they still may vary across the Galactic 
{Plane}. %Disk
Already within {our} toy models of the CR population (Model A and B), such variability can arise because of variations of the relative importance of different source populations within {the} Model A framework, or because of changes of the structure of Galactic magnetic field within the Model B framework. 

In the energy range above 100~TeV predictions of both models are inconsistent with LHAASO \gr\ data by a large margin. We have checked that this inconsistency is not 
{mitigated} %removed 
by changing the CR model setup (Model A vs.~Model B) 
{nor} %. We have also checked that it is not removed 
by changing the normalization of the diffuse flux at lower energies, within the margin allowed by Fermi-LAT modeling of the diffuse \gr\ emission from the Milky Way~
%disk 
\cite{2012ApJ...750....3A,Fermi-LAT:2016zaq}. We think that the conclusion about the discrepancy between the pion decay model and LHAASO data is robust. {Our results are consistent with earlier conclusions of Ref. \cite{Prevotat:2024ynu}. Similar conclusions have recently been reached in an independent study of Ref. \cite{Castro:2025wgf}.
}

{This can be {a} first indication that  
{the sources of CRs contributing to the knee are rare.} % at knee sources of cosmic rays are rare. 
Simulations of 
%contributions of rare 
{Galactic CR source populations} %sources in CR flux at knee in galaxy 
show that 
{CRs in the knee region quickly} 
escape 
{perpendicular to the disk} %out of disk 
before they can contribute to {the local} diffuse flux~\cite{Giacinti:2023upw}. As {a} result, {the} CR flux in {the} Galaxy {becomes} very non-uniform because PeV cosmic rays are present only in isolated bubbles, rather than a halo around the Galactic Disk.  In future, such {a} model can be tested {via} detailed $\gamma$-ray observations by LHAASO. It predicts strong variation of 100~TeV \gr\ flux along {the} Galactic Plane.} This is consistent with recent results of HAWC \cite{HAWC:2023wdq} and LHAASO \cite{Cao2025}, which found significant variation of flux level and power law index  along Galactic plane. 

In such a scenario, the locally observed knee of the CR spectrum is a   transient feature appearing in the CR spectra at different locations in the Galaxy following injection of ``fresh'' CRs by an individual source (for example, a supernova explosion) or a group of sources (for example, following an increased star formation activity episode)~\cite{Erlykin:1999fh,Bouyahiaoui:2018lew}. In this case, detection of the knee in the spectrum of CRs at the position of the Solar system only indicates that the Solar System is situated in one of the regions in which an episode injection of PeV energy CRs occurred recently.

Within our Model A, we can identify the contribution of this episode to the overall CR flux  with the highest energy (the 3rd) Peters cycle. This contribution is not generically present in other parts of the Galactic 
{Plane} %Disk 
and hence the pion decay emission from CRs contributing to the 3rd Peters cycle should not be considered as a part of the overall pion decay \gr\ flux. Model prediction for the pion decay \gr\ flux without account of the highest energy Peters cycle is shown by dashed line in Fig.~\ref{fig:gamma_flux}. One can see that prediction of this model does not exceed the LHAASO flux measurements above 100~TeV neither in the Inner nor in the Outer Galaxy. The overall pion decay flux from the Galactic
{Plane} %Disk
is a sum of the steady state flux from the bulk of the Galactic CR population (represented by Peters cycles 1 and 2 in our Model A) and of the transient diffuse emission from the local PeV CR bubble produced by CRs injected by the event responsible for the locally observed knee of the CR spectrum (the 3rd, Peters cycle in Model A). 

In the first approximation, the diffuse emission from the local interstellar medium can be calculated assuming that the CRs with the spectrum corresponding tot he 3rd Peters cycle in our Model A occupy a limited volume 
%bubble 
of radius $R$.  {The} solid red line in Fig.~\ref{fig:gamma_flux} shows a calculation of the pion decay flux 
{which} % with
account of emission of the local PeV CR bubble. We have adjusted the bubble radius in such a way that the sum of the steady-state diffuse emission from CR populations 
{representing} %on 
the 1st and 2nd Peters cycles and the transient emission from CRs forming the 3rd cycle and residing in the bubble fits LHAASO data. We find that a model with a bubble extending to the distance $R=0.7\pm 0.3$~kpc in the Outer Galaxy direction and up to $1.5\pm 0.3$~kpc in the Inner Galaxy direction is consistent with LHAASO diffuse flux measurements. These estimates are done within a ``toy model" of patchy distribution of PeV CRs in the Galactic disk. We expect that deeper exposure by LHAASO will provide more detailed information on the local PeV CR bubble geometry and on the overall distribution of PeV CR bubbles across the Galaxy.

\appendix
\section{Model based on Peters cycles}
We give in this section a few additional details about the way this model is computed. 

CR experiments use different technics to calibrate their energy, resulting in offsets between them. To perform the fit, energies of Auger data points were increased by $10\%$ so that it is more consistent with LHAASO’s data (we used IceTop data points to relate the two datasets). 

LHAASO and Auger divide the CR elements in four elemental groups: p, He, CNO and heavy. In order to remain consistent, we gathered AMS-02 measurements in similar groups, splitting measurements of intermediate elements (such as magnesium or aluminum) between the CNO and heavy groups based on their charge.
Also, DAMPE having measured only carbon, nitrogen and oxygen in the CNO group, we rescaled this spectrum to increase the intensity, so that the low energy part matches AMS-02’s measurements for which more elements were measured.
Finally, lowest energy points (corresponding to rigidities below $80 \; \mathrm{GV}$) were excluded from the fit, because the CR spectrum at the lowest energies is highly affected by solar modulations.

Best fit parameters of the Peters cycles model are presented in Tab. \ref{tab1 Peters} and \ref{tab2 Peters}.

\begin{table}[]
\centering
\begin{tabular}{|p{2.0cm}||p{1.3cm}p{1.5cm}p{1.2cm}|}
     \hline
     Parameters & $\mathcal{N}_1$ & $\mathcal{N}_2$ & $\mathcal{N}_3$ \\
     \hline
      Protons & 8.46e-2 & 1.38e-1 & 1.24e+0  \\
      Helium & 6.02e-2 & 1.51e-1 & 6.82-1 \\
      Intermediate & 4.14e-2 & 1.37e-1 & 4.64e-1 \\
      Heavy & 2.82e-2 & 1.00e-2 & 1.69e-1 \\
      \hline
\end{tabular}
\caption{Values of the normalisation constants $\mathcal{N}_i$ in Model A, in $(\mathrm{cm^2\,s\,sr})^{-1}\mathrm{GeV^{1.5}}$ (Eq.~\ref{eq:power_law}).}
\label{tab1 Peters}
\end{table}

\begin{table*}[]
\centering
\begin{tabular}{|p{1.7cm}||p{1cm}p{1cm}p{1cm}p{1cm}p{1cm}p{1cm}p{1cm}p{1cm}p{0.8cm}p{0.8cm}p{0.8cm}p{0.8cm}p{0.7cm}|}
     \hline
      Parameters  & $ {\cal R}_{01} $ & ${\cal R}_{10}$ & ${\cal R}_{11}$ & ${\cal R}_{20}$ & ${\cal R}_{21}$ & ${\cal R}_{30}$ & ${\cal R}_b$ & ${\cal R}_{31}$ & $\alpha_1$ & $\alpha_2$ & $\alpha_3$ & $\alpha_4$ & $\delta$ \\
     \hline
      Values  & 1.28e3 & 1.29e2 & 1.09e4 & 2.22e3 & 4.23e5 & 4.15e5 & 5.78e5 & 7.78e7 & 2.22 & 2.46 & 0.12 & 3.79 & 0.78 \\
      \hline

\end{tabular}
\caption{Rigidity thresholds ${\cal R}_{i0},{\cal R}_{i1}$ (in GV) and exponents $\alpha_i$  of the power laws used for the description of the CR energy spectra (Eq.~\ref{eq:power_law}) in Model A. ${\cal R}_{01}$ is the energy cut-off of the model used to account for solar modulations.}
\label{tab2 Peters}
\end{table*}

\section{DRAGON-2 simulation}
\label{Appendix:dragon}
The propagation of CRs can be described by a Boltzmann transport equation, which we solve numerically using the DRAGON-2 code~\cite{Evoli_2017}. The solution is obtained in a cylindrical symmetric coordinate system, represented in a 3D matrix, assuming the galactic box with galactocentric radius distances $r \in(0,20 \, \mathrm{kpc})$ and halo size $z \in (-4,4 \, \mathrm{kpc})$. For the set up of the simulations, we use a source distribution model based on Ref.~\cite{Ferri_re_2001}. The details of the source profile function are provided in 
Ref.~\cite{Evoli_2017}.

The diffusive CRs can be characterised by a diffusion coefficient $D(\mathcal{R})$ which can depend on both spatial coordinates and rigidity. For the simulation, we assume that the spatial diffusion coefficient is constant and homogeneous. In Eq.~\ref{eq: diffusion 5break}, the value of the parameters used in the first 3 breaks is adopted from Ref.~\cite{Schwefer_2023}. The parameters of the last 2 breaks are made to fit the LHAASO proton data \cite{LHAASO:2025byy}. The parameters are shown in Table \ref{parameter table}. 
\vspace*{1em}
\begin{table}[H]
\centering
\captionsetup{justification=centering}
\begin{tabular}{ll|l}
\hline
$\gamma_0$ &= 2.38 & Injection source spectral index \\
$D_0$ &= 5.18 & Normalization constant $10^{28}$ cm$^2$ s$^{-1}$ \\
$\delta_1$ &= 0.0116 & Diffusion spectra index \\
$\mathcal{R}_{12}$ &= 5.14 & First spectral break point rigidity (GV) \\
$s_{12}$ & = 0.063 & Softness parameter at first break\\
$\delta_2$ & = 0.566 & Second diffusion spectra index \\
$s_{23}$ &= 0.75 & Softness parameter at second break\\
$\mathcal{R}_{23}$ &= 372 & Second spectral break point rigidity (GV) \\
$\delta_3$ &= 0.159 & Third diffusion spectra index \\
$s_{34}$ &= 0.167 & Softness parameter at third break\\
$\mathcal{R}_{34}$ &= 10982 & Third spectra break point rigidity (GV) \\
$\delta_4$ &= 0.453 & Fourth diffusion spectra index \\
$s_{45}$ &= 0.022 & Softness parameter at fourth break\\
$\mathcal{R}_{45}$ &= 100000 & Fourth spectra break point rigidity (GV) \\
$\delta_5$ &= 0.15 & Fifth diffusion spectra index \\
$s_{56}$ &= 0.31 & Softness parameter at fifth break\\
$\mathcal{R}_{56}$ &= 4000000 & Fifth spectra break point rigidity (GV) \\
$\delta_6$ &= 1.5 &Sixth diffusion spectra index \\
\hline
\end{tabular}
\caption{Parameter table for the spectral break.}
\label{parameter table}
\end{table}

The CRs interact with the gas in the interstellar medium (ISM), which is mainly composed of hydrogen in various states, including atomic (H$_\mathrm{I}$), molecular (H$_2$), and ionised hydrogen (H$_\mathrm{II})$, resulting in the production of secondary particles. The number density of the hydrogen with spatial dependence is calculated as,
\begin{equation}
    n(r,z) = n_\mathrm{H_I}+2n_\mathrm{H_2}+n_\mathrm{H_{II}}.
\end{equation}
The simulation uses gas models of $\mathrm{H_{I}}$, $\mathrm{H_2}$ and $\mathrm{H_{II}}$ based on Refs.~\cite{Gordon_1976, Bronfman_1988, Cordes_1991}, respectively.

\section{Gamma-ray absorption model}

\begin{figure}
    \centering
    \includegraphics[width=1\linewidth]{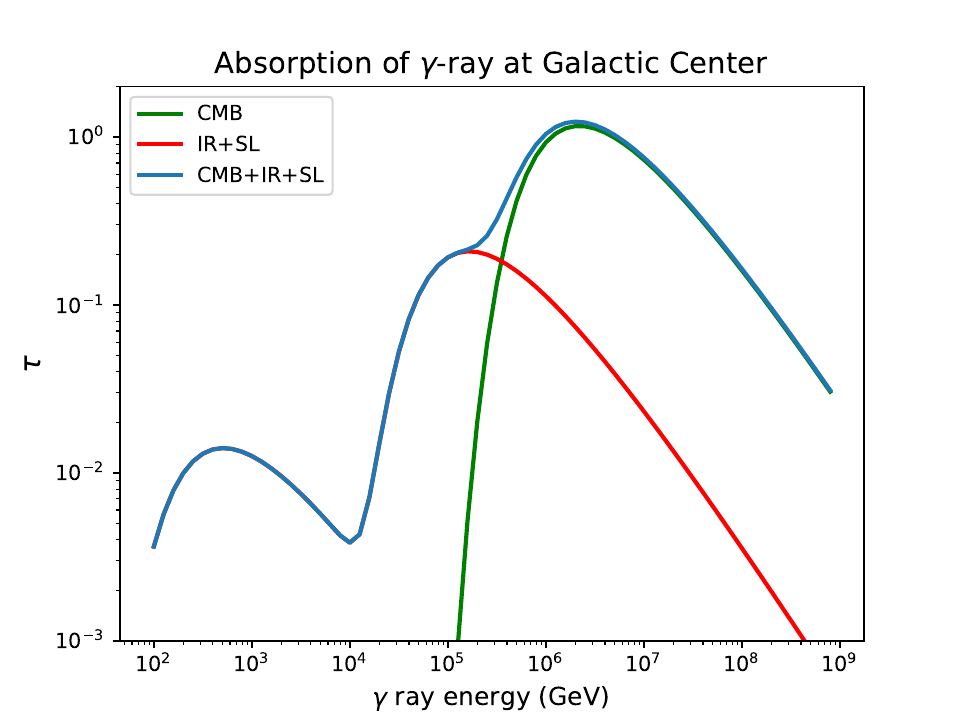}
    \caption{Absorption of the $\gamma$-ray at the galactic center.}
    \label{fig:absorption at GC}
\end{figure}

The $\gamma$-ray with energy $\geq 30$ TeV~\cite{Vernetto2016} experiences a non-negligible absorption effect due to the pair production interaction $(\gamma  + \gamma \rightarrow e^+ + e^-)$. The photon density in the galaxy can be characterized into 3 types. The largest and primary one is the cosmic microwave background (CMB), which is isotropic and uniformly distributed in the whole space. The other two components considered are starlight (SL) and infrared (IR), both of which originate from stellar and dust emission within the Galaxy. The total photon density $n_\gamma$ can be expressed by the sum of three components,
\begin{equation}
    n_\gamma = n_{CMB}+n_{SL}+n_{IR}.
\end{equation}
 The probability that a photon with energy $E$ will reach Earth without being absorbed is governed by the optical depth $\tau$ associated with these photon fields.
\begin{equation}
    P(E) = e^{-\tau(E)},
\end{equation}
For different radiation fields, we calculate the opacity based on Ref.~\cite{Celli_2017} with the form,
\begin{equation} \label{eq:opacity master}
    \tau_i = \frac{1}{\pi} r_e^2 L_i T^3_i \xi_i f\left({m_e^2}/{T_i E_\gamma}\right) \,,
\end{equation}
where the $r_e$ is classical electron radius, $L_i$ is the effective size of the radiation field, $T_i$ is temperature, $\xi_i$ is the non-thermic parameter and function $f(x) = -a x\ln[1-\exp(-bx^c)]$, where $a=3.68$, $b=1.52$ and $c=0.89$. The effective size of SL and IR at a given galactic location,  as experienced by a photon traveling to the Sun's position, can be calculated by their density $\rho_i(r,z)$,
\begin{equation}
    L_i(r,z) = \int dl \frac{\rho_i(r,z)}{\rho_i^{GC}},
\end{equation}
% \MA{\bf (What is the relation between $l$, $r$ and $z$ in this integral?)}
where $\rho_i^{GC}$ is density at the galactic center, $r$ and $z$ are the cylindrical coordinates, $dl$ is the step size with a value $0.1$~kpc. The density can be further expressed based on the models of Ref.~\cite{Vernetto2016}. For SL,
\begin{equation}
    \rho_{SL}(r,z) = \rho_{SL}^{GC} \;\text{exp}\left(\frac{-r}{R_{SL}}-\frac{|z|}{Z_{SL}}\right)\,.
\end{equation}
For IR,
\begin{equation} 
    \rho_{c,w}(r,z) = \rho^{GC}_{c,w} \text{exp} \left(\frac{-r}{R_{c,w}}-\frac{|z|}{Z_{c,w}}\right)\,,
\end{equation}
where the subscript $c$ and $w$ denotes cold and warm dust respectively. The temperature of the cold dust has a spatial dependence given by,
\begin{equation}
    T_c(r,z) = (T_c^{GC} - T_\infty) \; \text{exp} \left(\frac{-r}{R_T}-\frac{|z|}{Z_T}\right) + T_\infty\,.
\end{equation}
The value of the parameters adapted from Ref.~\cite{Vernetto2016} is shown in Table~\ref{parameter table for absorption}. The opacity of different radiation fields at the galactic center is shown in Figure~\ref{fig:absorption at GC}.
\vspace*{1em}
\begin{table}[H]
\centering
\captionsetup{justification=centering}
\begin{tabular}{lll}
\hline
Parameter & Value & Units \\
\hline
$\rho^{GC}_{c}$  & $1.51\times10^{-25}$  & g\;cm$^{-3}$ \\
$\rho^{GC}_{w}$  & $1.22\times10^{-27}$  & g\;cm$^{-3}$ \\
$R_{SL}$         & 2.17 (r $<8$)         & kpc          \\
$R_{SL}$         & 4.07 (r $\geq8$)      & kpc          \\
$Z_{SL}$         & 7.22                  & kpc          \\
$R_{c}$          & 5                     & kpc          \\
$R_{w}$          & 3.3                   & kpc          \\
$Z_{c}$          & 0.1                   & kpc          \\
$Z_{w}$          & 0.09                  & kpc          \\
$R_{T}$          & 48                    & kpc          \\
$Z_{T}$          & 500                   & kpc          \\
$T_{w}$          & 35                    & K            \\
$T^{GC}_{c}$     & 19.2                  & K            \\
$T_{\infty}$     & 2.7                   & K            \\
\hline
\end{tabular}
\caption{Parameter table for $\gamma$-ray absorption model.}
\label{parameter table for absorption}
\end{table}

%\begin{figure}
%    \centering
 %   \includegraphics[width=1\linewidth]{CR_M1_vs_M2_by_comp.pdf}
 %   \caption{Fig. for internal discussion. M1 -- Clement (full line) vs M2 -- DRAGON (dashed) CR spectrum by components for solar system. Dots are AMS-02 data.}
 %   \label{fig:CR_M1_vs_M2}
%\end{figure}

%\begin{figure}
%    \centering
%    \includegraphics[width=1\linewidth]{gamma_M1_vs_M2_by_comp.pdf}
%    \caption{Fig. for internal discussion. Gamma-ray emissivity for figure \ref{fig:CR_M1_vs_M2}. The legend is the same (with orange p+He and black p+He+CNO).}
%    \label{fig:gam_M1_vs_M2}
%\end{figure}

\bibliography{refs.bib}

\end{document}